\documentclass[a4paper]{aa}
\usepackage{psfig,times}

\psfigurepath{/z/rue/couette:/z/rue/plots/schultzplot}

\begin{document}
\def\apj{{ApJ}}       
\def\aj{{AJ}}       
\def\apjs{{ Ap. J. Suppl.}} 
\def\apjl{{ Ap. J. Letters}} 
\def\pasp{{ Pub. A.S.P.}} 
\def\mn{{MNRAS}} 
\def\aa{{A\&A}} 
\def\aasup{{ Astr. Ap. Suppl.}} 
\def\baas{{ Bull. A.A.S.}} 
\def\csss{{Cool Stars, Stellar Systems, and the Sun}\ }
\def\an{{Astron. Nachr.}}
\def\sp{{Solar Phys.}}
\def\gafd{{Geophys. Astrophys. Fluid Dyn.}}
\def\ass{{Ap\&SS}}
\def\acta{{Acta Astron.}}
\def\jfm{{J. Fluid Mech.}}
\def\ea{{et\thinspace al.\ }}   
\def\AIP{Astrophysikalisches Institut Potsdam}
\def\gR{G. R\"udiger}
\def\R{R\"udiger}
\def\lK{L.L. Kitchatinov}
\def\K{Kitchatinov}
\def\mK{{M.~K\"uker}\ }
\def\S{St\c{e}pie\`{n}}
\def\DR{differential rotation\ }
\def\qq{\qquad\qquad}                      
\def\qqq{\qquad\qquad\qquad}               
\def\q{\qquad}
\def\beg{\begin{eqnarray}}
\def\ende{\end{eqnarray}}
\def\Oms{ {\Omega^*} }
\def\O2{ {\delta \tilde{\Omega} }}
\def\Om{{\it \Omega}}
\def\bib{\item}
\def\lsim{\lower.4ex\hbox{$\;\buildrel <\over{\scriptstyle\sim}\;$}}
\newcommand{\Ha}{\mbox{Ha}}
\newcommand{\Pm}{\mbox{Pm}}

\title{Nonaxisymmetric patterns in the linear theory of MHD Taylor-Couette instability}
\author{D.A. Shalybkov\inst{1,2}
        \and\  G. R\"udiger\inst{1} \and M. Schultz\inst{1}}
\offprints{gruediger@aip.de}
\institute{ 
Astrophysikalisches Institut Potsdam,
An der Sternwarte 16, D-14482 Potsdam, Germany \and A.F. Ioffe Institute of 
Physics and Technology, 194021 St. Petersburg,
Russia}
\date{\today}
\abstract{The linear stability of MHD Taylor-Couette flow of infinite
vertical extension  is considered for various magnetic Prandtl numbers Pm.
The calculations are  performed for a wide gap container with 
$\hat\eta=0.5$ with an axial uniform magnetic field excluding counterrotating 
cylinders. For both hydrodynamically stable and unstable 
flows the magnetorotational instability
produces characteristic minima of the Reynolds number for
certain (low) magnetic field amplitudes and Pm $>$ 0.01.
For Pm $\lsim$ 1 there is a characteristic magnetic field amplitude
 beyond which  the
instability sets in in form of nonaxisymmetric spirals with
the azimuthal number $m=1$. Obviously, the magnetic field is able to excite nonaxisymmetric configurations despite of the tendency of differential rotation to favor axisymmetric magnetic fields which is known from the dynamo theory. If Pm is too big or too small, however,  the 
axisymmetric mode with $m$=0 appears to be the most  unstable one possessing  the lowest  Reynolds numbers --  as it is also true for hydrodynamic
Taylor-Couette flow or for very weak fields. That the most unstable mode
for modest Pm proves to be nonaxisymmetric must  be  considered as a strong indication for the
possibility of dynamo processes in connection with the magnetorotational
instability.     
\keywords{magnetohydrodynamics -- accretion disks -- turbulence}}
\titlerunning{Nonaxisymmetric patterns of the linear MHD Taylor-Couette instability}
\authorrunning{ D.A. Shalybkov \& G. R\"udiger \& M. Schultz}

\maketitle

\section{Introduction}
In order to discuss possible experimental realizations of the magnetorotational
instability as the main transporter of angular momentum in all kinds of
accretion disks there are several recent studies of Taylor-Couette flow for
electro-conducting fluids between rotating cylinders under the influence of an
uniform axial magnetic field (Ji et al. 2001; R\"udiger \& Zhang 2001; 
Willis \& Barenghi 2002). The numbers describing
the geometry of the container and the magnetic Prandtl number of the fluid have
been considered as the free parameters. For a given magnetic field amplitude (the Hartmann number) the critical 
angular velocity of the inner cylinder (the critical Reynolds number) is computed for the onset of an 
instability of the rotation law between the cylinders.

In R\"udiger \& Shalybkov (2002) the instability pattern is
considered as axisymmetric. The main result for resting outer cylinder is that
for high magnetic Prandtl number  for weak magnetic field the excitation of
the instability is easier than without magnetic field but for strong magnetic field the excitation of
the instability is more complicated. The effect, however,
disappears for small magnetic Prandtl number, i.e. for lower electric
conductivity of the fluid as it may be realized in protoplanetary disks.

On the other hand for rotating outer cylinder, when no instability without
magnetic field exists, the magnetic field always produces critical Reynolds
numbers which, however, are running with 1/Pm. For Pm of order $10^{-5}$ the
critical Reynolds number is of order $10^6$ which is just the experimental limit.
\begin{figure}
\psfig{figure=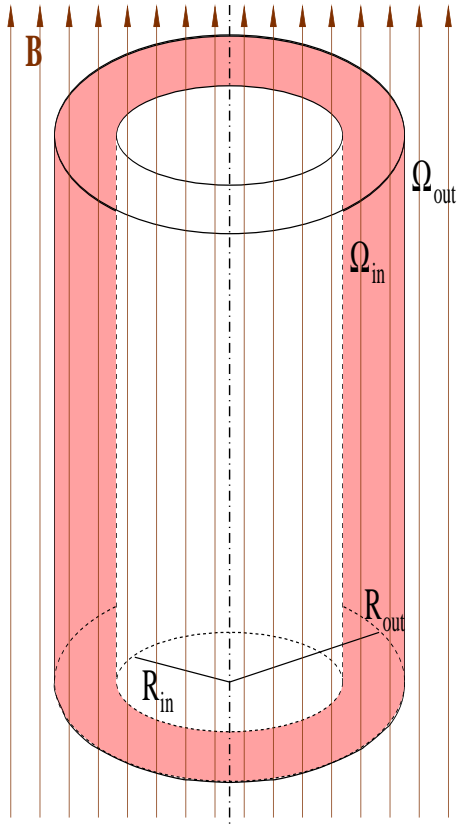,width=6cm,height=11cm}
\caption{\label{geometry} Cylinder geometry of the  Taylor-Couette flow}
\end{figure}

In the present paper the nonaxisymmetric perturbations are included into the
consideration. This is of particular relevance for the question whether the
Cowling theorem for dynamo action can be fulfilled, after which a dynamo can
only work with nonaxisymmetric fields. We shall find that indeed for certain
parameters  -- despite the smoothing action of the differential rotation --
nonaxisymmetric modes can be excited easier than axisymmetric modes. This is in
great contrast to earlier results of Taylor-Couette flow without magnetic fields
where always the axisymmetric modes possess the lowest Reynolds numbers
(Roberts 1965; DiPrima 1961)\footnote{For counterrotating cylinders, however, the preference of nonaxisymmetric modes is already known, see Kr\"uger et al. 1966; Chen \& Chang   1998 }.

Here,  the dependence of a real
Taylor-Couette flow on the magnetic Prandtl number and on the azimuthal
`quantum number $m$' is investigated.  The simple model of uniform  density fluid
contained between two vertically-infinite rotating cylinders is used with 
a constant magnetic field parallel to the rotation axis. The unperturbed state 
is a
stationary circular flow with $\Om$ 
\begin{equation}
\Om(r) = a+b/{R}^2,
\label{Om}
\end{equation}
where $a$ and $b$ are two constants related to the angular
velocities $\Om_{\rm in}$ and $\Om_{\rm out}$ with which the inner
and the outer cylinders are rotating. If $R_{\rm in}$ and $R_{\rm out}$
($R_{\rm out}>R_{\rm in}$) are the radii of the two cylinders then
\begin{equation}
a={\hat \mu-{\hat\eta}^2\over1-{\hat\eta}^2} \ \Om_{\rm in}, \quad\quad\quad 
b=  {1-\hat\mu\over1-{\hat\eta}^2}\ \Om_{\rm in} \ R_{\rm in}^2,
\label{ab}
\end{equation}
with 
\begin{equation}
\hat\mu=\Om_{\rm out}/\Om_{\rm in} \q {\rm and} \q
\hat\eta=R_{\rm in}/R_{\rm out}.
\label{mueta}
\end{equation}
After the Rayleigh stability criterion,
$d(R^2 \Om)^2/dR>0$,
rotation laws
are hydrodynamically stable
for $\hat\mu>\hat\eta^2$. Taylor-Couette flows with resting outer cylinders 
($\hat\mu=0$) are thus never stable. 
Here in order to isolate the MRI, we 
are also interested in flows with rotating outer cylinders  so that the hydrodynamical 
stability criterion,   $\hat\mu>\hat\eta^2$,  is fulfilled.
Our standard examples are formed with $\hat\eta=0.5$, $\hat\mu=0$ and 
$\hat\mu=0.33$, resp. The first example ($\hat\mu=0$) is hydrodynamically 
unstable 
%
%
and the second 
one  ($\hat\mu=0.33$) is hydrodynamically stable. We are here only 
interested  in flow patterns  in containers with positive $\hat \mu$.

\section{Basic equations}
The MHD equations which have to be solved are 
\beg
{\partial \vec{u} \over \partial t} + (\vec{u} \nabla)\vec{u} = - {1\over \rho}
\nabla p + \nu \Delta \vec{u} + \vec{J} \times \vec{B}
\label{0}
\ende
and
\beg
{\partial \vec{B} \over \partial t}= {\rm rot} (\vec{u} \times \vec{B}) + \eta \Delta
\vec{B}
\label{0.1}
\ende
with
$
{\rm div} \vec{u} = {\rm div} \vec{B} = 0$.
They are considered in cylindrical geometry with 
$R$, $\phi$, and $z$ as the cylindrical coordinates.  A viscous 
electrically-conducting incompressible fluid between
two rotating infinite cylinders in the presence of a uniform magnetic
field  parallel to the rotation axis leads to the basic solution
$U_R=U_z=B_R=B_\phi=0,
B_z=B_0={\rm const.,} \ {\rm and} \  \Om=a+b/R^2$, 
with  $\vec{U}$ as  the flow  and  $\vec{B}$ as the magnetic field. 
We are interested in the stability of
this solution. The perturbed state of the flow may be  described by
$u_R', \; U_\phi+u_\phi', \; u_z', \; B_R', \; B_\phi', \; B_0+B_z',
\; p',$
with  $p'$ as the  pressure perturbation.

Here only  the linear stability problem is considered.
By analyzing the disturbances into normal modes  the solutions
of the linearized magnetohydrodynamical equations are of the form
\begin{eqnarray}
\lefteqn{u_R'=u_R(R)e^{{\rm i}(m\phi+kz-\omega t)}, 
\ \ \  B_R'=B_R(R)e^{{\rm i}(m\phi+kz-\omega t)},}  \nonumber \\
\lefteqn{u_\phi'=u_\phi(R)e^{{\rm i}(m\phi+kz-\omega t)},
\ \ \ \   B_\phi'=B_\phi(R)e^{{\rm i}(m\phi+kz-\omega t)},} \nonumber \\
\lefteqn{u_z'=u_z(R)e^{{\rm i}(m\phi+kz-\omega t)},
\ \ \ \ \  B_z'=B_z(R)e^{{\rm i}(m\phi+kz-\omega t)}.}
\end{eqnarray}
 Only marginal stability  will be considered where the imaginary  part of $\omega$
 vanishes.
Let  $d=R_{\rm out} - R_{\rm in}$
be the gap between
the cylinders. We use 
\begin{equation}
H=(R_{\rm in}d)^{1/2}
\label{2.4}
\end{equation}
 as unit of length,
the  $ \eta /H$ as unit of
perturbed velocity and $B_0 $ as unit of perturbed
magnetic field with the magnetic Prandtl number 
\begin{equation}
{\rm Pm} = {\nu\over\eta},
\label{pm}
\end{equation}
 $\nu$ is the
kinematic viscosity, $\eta$ is the magnetic diffusivity. 
Note  $H^{-1}$ as the unit of wave numbers and $\nu/H^2$ as the unit of frequencies.
After elimination of both pressure fluctuations and the fluctuations of the vertical magnetic field, $B'_z$, the equations are
\beg
{\partial u_R \over \partial R} + {u_R \over R} + {{\rm i}m \over R} u_\phi + 
{\rm i}k u_z = 0,
\label{1}
\ende
\begin{eqnarray}
\lefteqn{{\partial^2 u_\phi \over \partial R^2} + {1\over R} {\partial u_\phi
\over \partial R} - {u_\phi \over R^2} - \left({m^2 \over R^2} + k^2\right)
u_\phi -}\nonumber\\
\lefteqn{-{\rm i} \left(m {\rm Re} {\Om \over \Om_{\rm in}} - \omega\right)
u_\phi + {2{\rm i}m \over R^2} u_R - {\rm Re} {1\over R} 
{\partial \over \partial R}
\left(R^2 {\Om \over \Om_{\rm in}}\right) u_R} \nonumber\\
&& - {m \over k} \left[{1\over R} {\partial^2 u_z \over \partial R^2} + 
{1\over R^2} {\partial u_z \over \partial R} - \left({m^2 \over R^2} + 
k^2\right) {u_z
\over R} - \right. \nonumber\\
&& \left. - {\rm i}\left(m {\rm Re} {\Om \over \Om_{\rm in}} - \omega\right) 
{u_z
\over R}\right]  + {m\over k} {\rm Ha}^2 \left[{1\over R} {\partial B_R \over 
\partial R} +
{B_R \over R^2}\right] +\nonumber\\
&& + {{\rm i}\over k} {\rm Ha}^2 \left({m^2\over R^2} + k^2\right)
B_\phi = 0,
\label{2}
\end{eqnarray}
\begin{eqnarray}
\lefteqn{{\partial^3 u_z \over \partial R^3} + {1\over R} 
{\partial^2 u_z \over \partial
R^2} - {1\over R^2} {\partial u_z \over \partial R} - \left({m^2\over R^2} +
k^2\right) {\partial u_z \over \partial R} +}\nonumber\\
\lefteqn{+{2m^2 \over R^3} u_z - {\rm i}\left(m
{\rm Re} {\Om\over \Om_ {\rm in}} - \omega\right) {\partial u_z \over
\partial R} -  
 {\rm i}m {\rm Re} {\partial \over \partial R} \left({\Om \over \Om_{\rm
in}}\right) u_z} \nonumber\\ 
&& - {\rm Ha}^2 \left[{\partial^2 B_R \over \partial R^2} + {1\over
R} {\partial B_R \over \partial R} - {B_R \over R^2} - k^2 B_R + \right. \nonumber\\
&& \left. +{{\rm i}m \over R}
{\partial B_\phi \over \partial R} - {{\rm i}m\over R^2} B_\phi \right] -
{\rm i}k\left[{\partial^2 u_R \over \partial R^2} + {1\over R} {\partial u_R
\over \partial R} - {u_R \over R^2} - \right. \nonumber\\
&& \left. - \left(k^2 + {m^2\over R^2}\right)
u_R\right] - k \left(m {\rm Re} {\Om \over \Om_{\rm in}} - \omega\right)
u_R -\nonumber\\
&& - 2 {km \over R^2} u_\phi - 2 {\rm i}k {\rm Re} {\Om \over \Om_{\rm in}}
u_\phi = 0,
\label{3}
\end{eqnarray}
\begin{eqnarray}
\lefteqn{{\partial^2 B_R \over \partial R^2} + {1\over R} {\partial B_R \over 
\partial R}
- {B_R \over R^2} - \left({m^2 \over R^2} + k^2\right) B_R -}\nonumber\\
&& - {2{\rm i}m\over R^2}
B_\phi - {\rm i} {\rm Pm} \left(m {\rm Re} {\Om \over \Om_{\rm in}}
-\omega\right) B_R + {\rm i}k u_R=0,
\label{4}
\end{eqnarray}
\begin{eqnarray}
\lefteqn{{\partial^2 B_\phi \over \partial R^2} + {1\over R} 
{\partial B_\phi \over
\partial R} -{B_\phi \over R^2} - \left({m^2 \over R^2} + k^2\right) B_\phi
+}\nonumber\\
&& +{2{\rm i}m \over R^2} B_R -{\rm i} {\rm Pm} \left(m {\rm Re} 
{\Om \over \Om_{\rm in}}
- \omega\right) B_\phi + {\rm i}k u_\phi+\nonumber\\ 
&& + {\rm Pm} \ {\rm Re} \ R {\partial \Om
/\Om_{\rm in} \over \partial R} B_R = 0.
\label{5}
\end{eqnarray}
The Reynolds number Re and the Hartmann number Ha are defined as
\beg
 {\rm Re} = {\Om_{\rm
in} H^2 \over \nu}, \qqq {\rm Ha} = {B_0 H \over \sqrt{\mu_0 \rho \nu \eta}}.
\label{HaRe}
\ende
For  given Hartmann number and Prandtl number in the present paper we shall derive in a linear theory the critical Reynolds number of the rotation of the inner cylinder for various mode numbers $m$.
\section{Boundary conditions, numerics}
An appropriate set of ten boundary conditions is needed to solve  the system
(\ref{1})..(\ref{5}). 
Always no-slip conditions for the velocity on the walls
are used, i.e. 
$
u_R=u_\phi=d u_R/dR=0.$
The boundary conditions depend on the electrical properties
of the walls. The transverse currents and the perpendicular component of the 
magnetic field  vanish on conducting walls  hence
$
d B_\phi/dR + B_\phi/R = B_R = 0.$
These   boundary conditions  hold  both 
for $R=R_{\rm in}$ and  for $R=R_{\rm out}$.

\begin{figure}
\psfig{figure=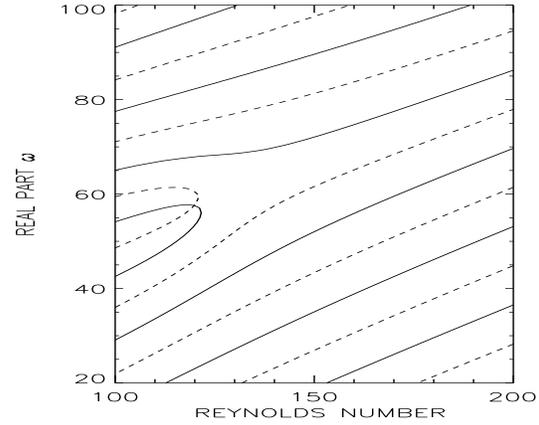,width=8cm,height=6cm}
\caption{\label{f00} The zero-lines of the real (solid) and the imaginary (dashed) part of the system determinant. At the crossing point both parts of the determinant simultaneously vanish fixing the Reynolds number and the drift frequency $\cal{R}(\omega)$. Pm and $k_z$ are prescribed}
\end{figure}

The homogeneous set of equations (\ref{1})...(\ref{5}) with the boundary
conditions  for conducting walls  determine the eigenvalue problem of the form 
$
{\cal L}(k, m,  {\rm Re}, {\rm Ha}, {\cal {R}}(\omega))=0
$ 
for given Pm. The real part of $\omega$,  ${\cal {R}}(\omega)$, describes a  
drift of the pattern along the azimuth which only exists for nonaxisymmetric 
flows. $\cal L$ is a complex quantity, both its real part and its imaginary part 
must vanish for the critical Reynolds number (Fig. \ref{f00}). The latter is 
minimized by choice of the wave number $k$. ${\cal R}(\omega)$ is the second
quantity which is fixed by the eigenequation.    

The system is approximated by finite differences with typically 81 gridpoints. The
resulting determinant, $\cal L$, takes the value zero if and only if the 
values  Re  are the eigenvalues.
For a fixed Hartmann number, a fixed Prandtl number and a given vertical wave 
number we find the eigenvalues of the equation system. They are always minimal 
for a certain wave number which by itself defines the marginally unstable mode.
The corresponding eigenvalue is the desired Reynolds number.

\section{Results for conducting walls}
Only a  container is considered in the present paper with one and the 
same gap geometry, i.e. $\hat \eta = 0.5$. Then the  flow between the 
cylinders is hydrodynamically unstable   between $\hat \mu =0$ and 
$\hat \mu =0.25$. We shall work with both the hydrodynamically unstable
container with $\hat\mu = 0$ and with the hydrodynamically stable container
with $\hat\mu = 0.33$.
\subsection{Resting outer cylinder  (steep rotation law)}
\begin{figure}[ht]
\vbox{
\hbox{
\psfig{figure=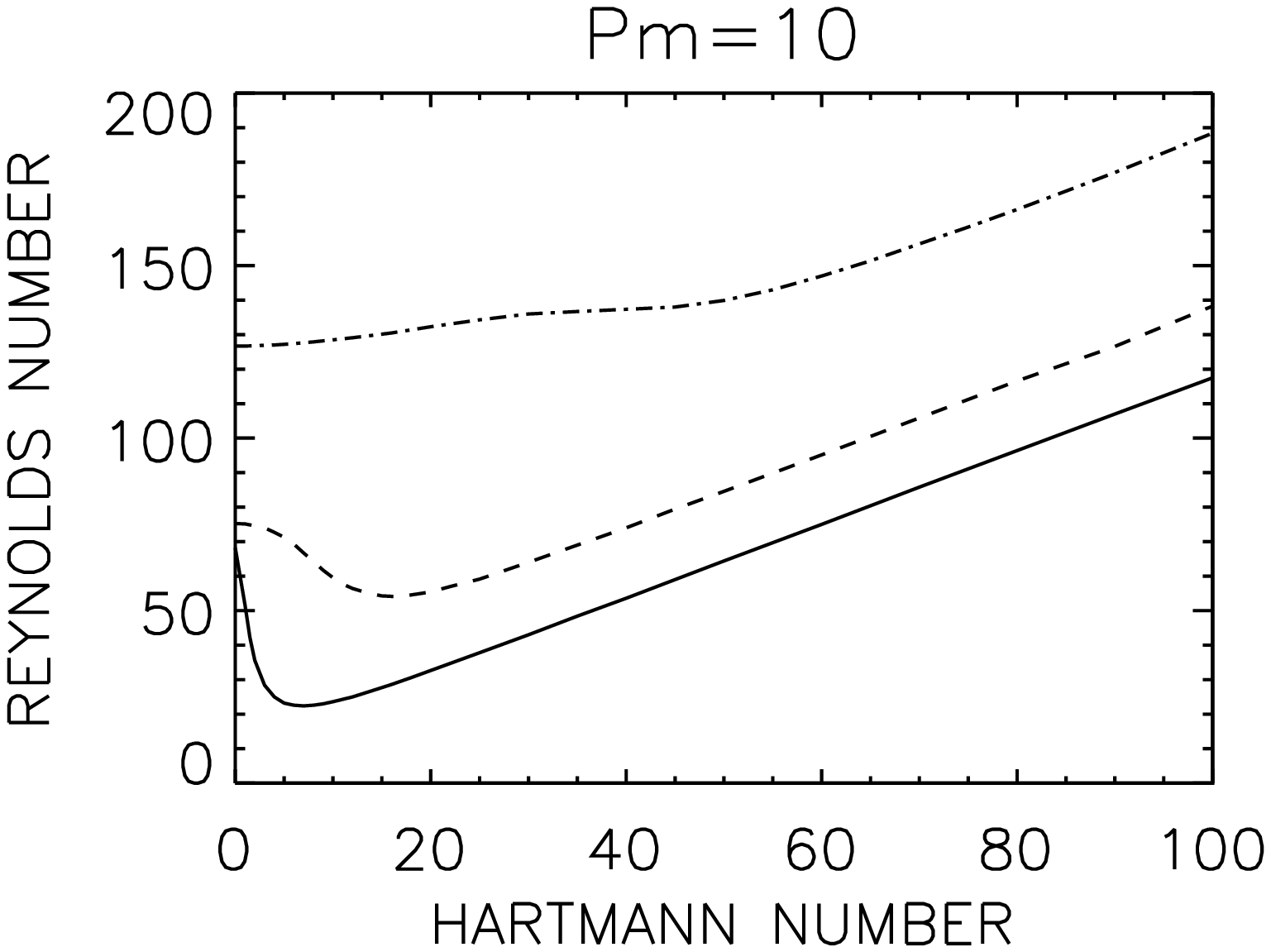,width=4cm,height=8cm}\hfill
\psfig{figure=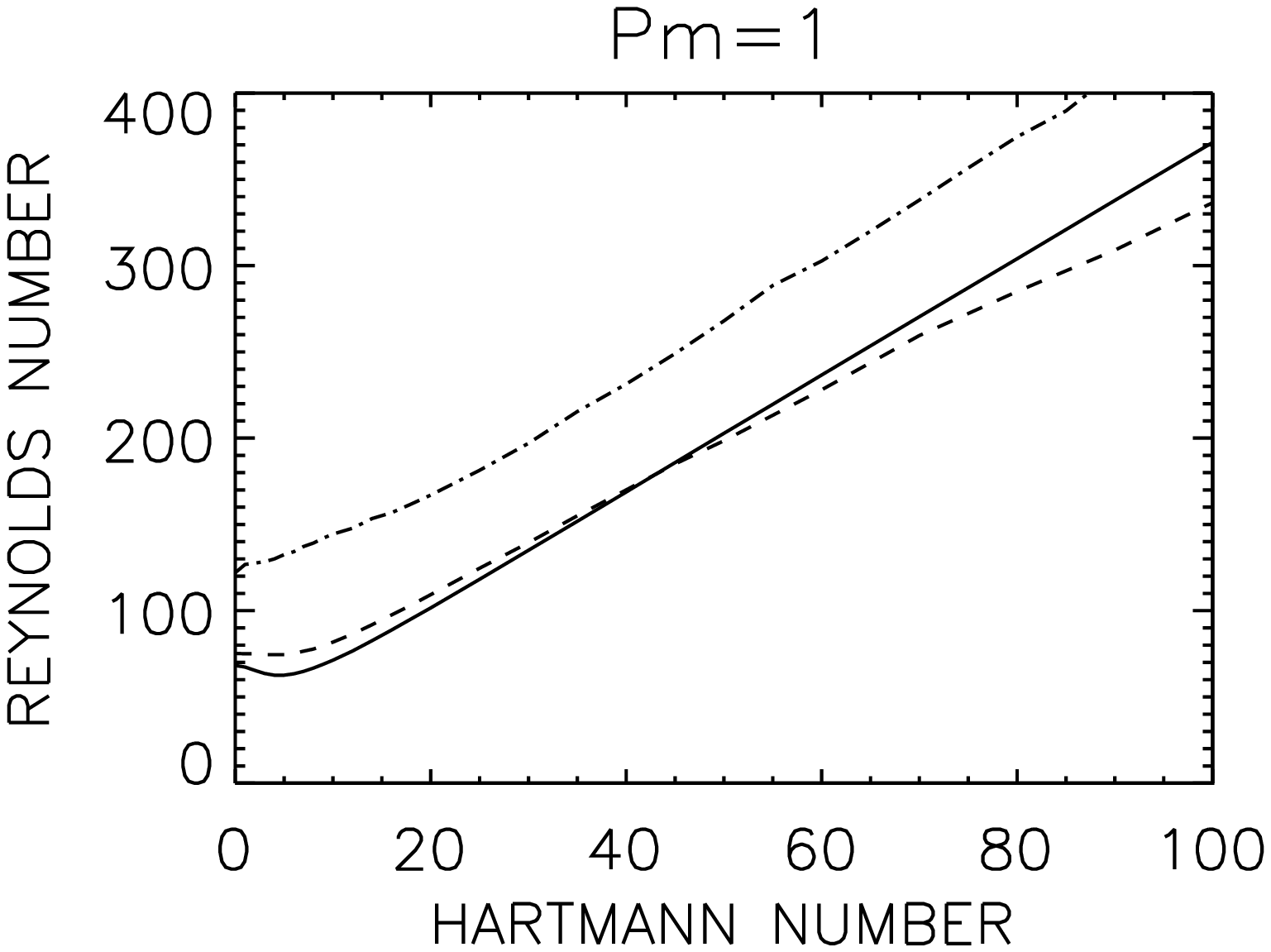,width=4cm,height=8cm}}
\hbox{
\psfig{figure=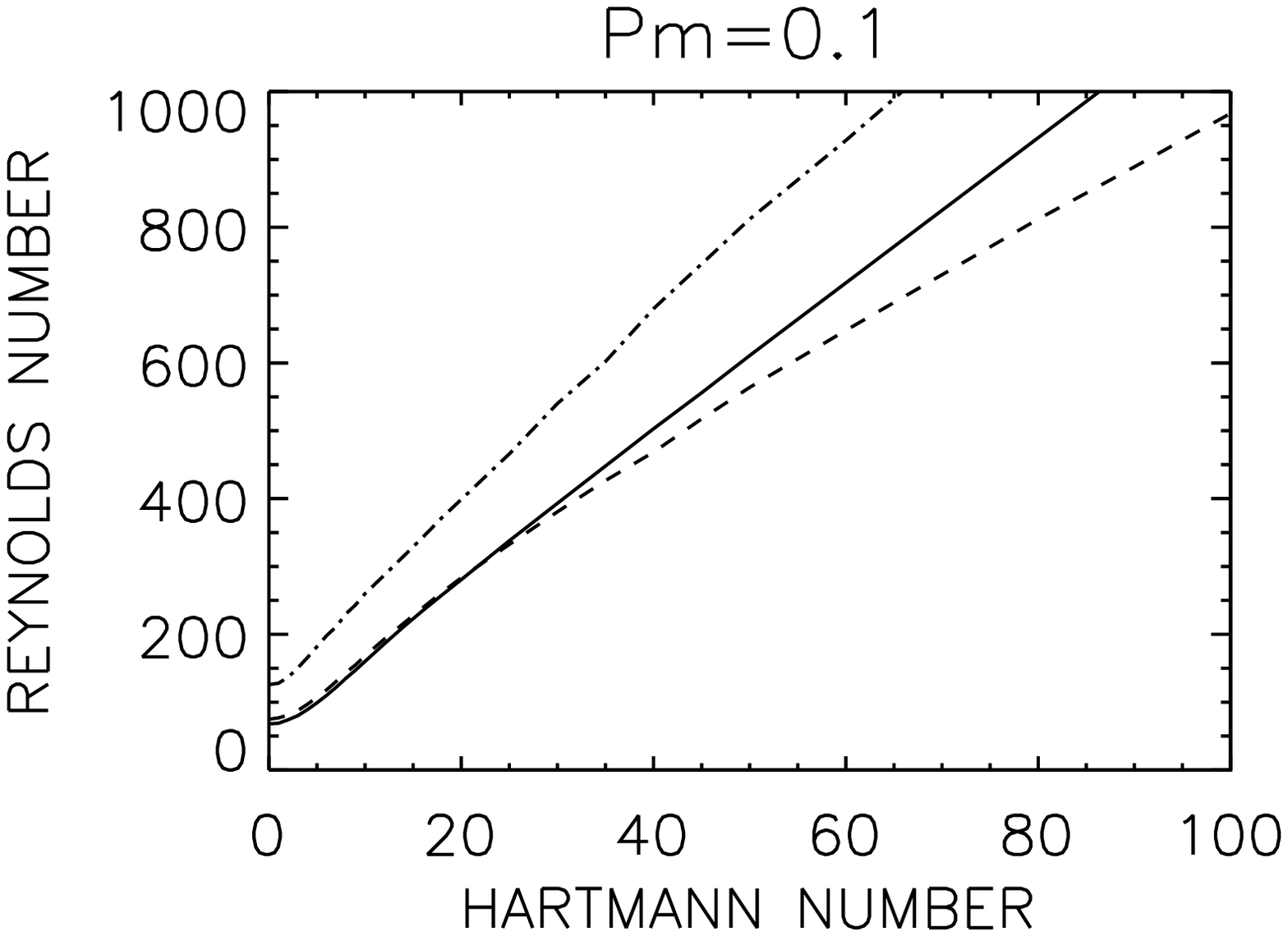,width=4cm,height=8cm}\hfill
\psfig{figure=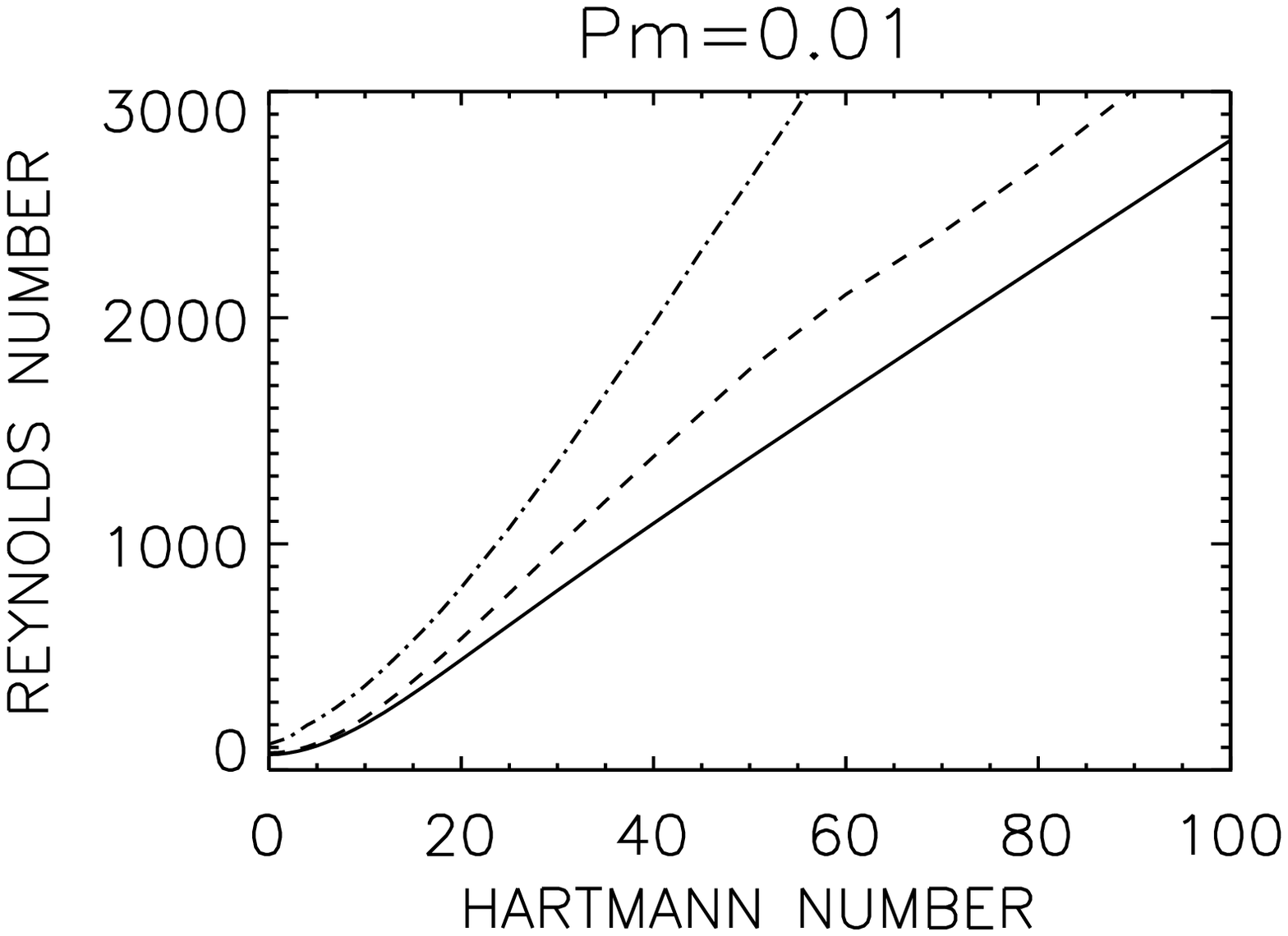,width=4cm,height=8cm}}}
\caption{\label{f1} Resting outer cylinder: Stability lines for axisymmetric ($m=0$, solid lines)
and nonaxisymmetric instability modes ($m=1$ (dashed lines), $m=2$ (dashed-dotted lines).  Results are given for  Pm $=10, 1, 
0.1$ and 0.01. Note that for Pm $=1$ and for Pm $=0.1$ for certain 
magnetic fields the nonaxisymmetric modes with $m=1$  possess the lowest Reynolds 
numbers}
\end{figure}

We start with the results for  containers  with  resting 
outer cylinders (Fig. \ref{f1}). Provided a critical  rotation rate of the inner cylinder is exceeded they  are hydrodynamically unstable. Of course, for Ha $=m=0$ the known critical Reynolds
number Re $=68$ is reproduced. For $m>0$ the critical Reynolds numbers exceed the
value for $m=0$. Without magnetic field the instability yields rolls. The critical Reynolds number for $m=1$ is   73  and for $m=2$ it is   101 (Roberts 1965).

With magnetic fields (Ha $>$0) the magnetic Prandtl number comes into the game.
Results for Pm $=10, 1, 0.1$ and 0.01 are presented  in Fig. \ref{f1}. For
Pm $\geq 1$ the electrical conductivity is so high that the magnetorotational
instability (Balbus \& Hawley 1991; Brandenburg et al. 1995; Ziegler \& 
R\"udiger 2000) for Ha $\simeq 5$
produces a characteristic minimum of the critical Reynolds number but for
stronger magnetic fields the suppressing action of the magnetic field starts to
dominate. In contrast to the expectations, however, for the magnetic  Prandtl numbers which are not too high and not too low   the
mode with $m=1$ becomes more and more dominant. This is a new and interesting
result: The linear instability of the Taylor-Couette flow without magnetic field
is formed by axisymmetric rolls but the magnetic field favors the excitation of
bisymmetric spirals. For Ha $>10\dots 20$ the instability sets in in form of a
drifting pattern with maximum and minimum separated by 180$^\circ$.  However, as can be seen in Fig. \ref{f1} (last plot) for small magnetic Prandtl number (here Pm=0.01) again the axisymmetric pattern with $m=0$ again starts to dominate with the lowest critical Reynolds number.

The modes with $m=2$,  which we have also computed,  do never possess the lowest Reynolds numbers, they are not important for the discussion of the pattern of  the instability. What we have found is that in contrast to the hydrodynamic case (Pm=0) there are experimental combinations where the nonaxisymmetric mode with $m=1$ has a lower Reynolds number than the axisymmetric mode with $m=0$. This is one of the most surprising structure-forming consequences of the inclusion of magnetic fields to the Taylor-Couette flow experiment found first in astrophysical simulations. 
\subsection{Rotating outer cylinder  (flat rotation law)}
\begin{figure}[ht]
\vbox{
\hbox{
\psfig{figure=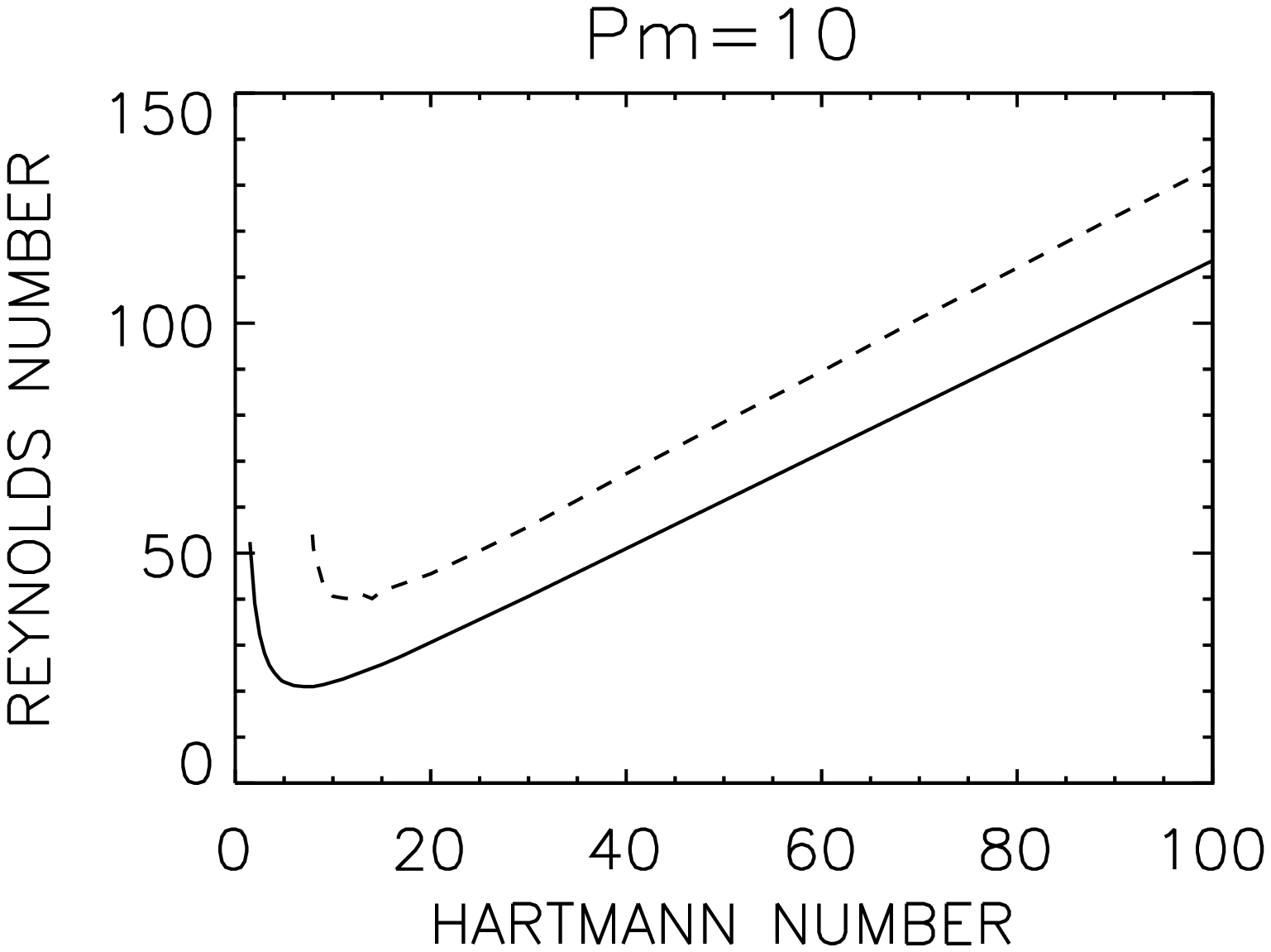,width=4cm,height=8cm}\hfill
\psfig{figure=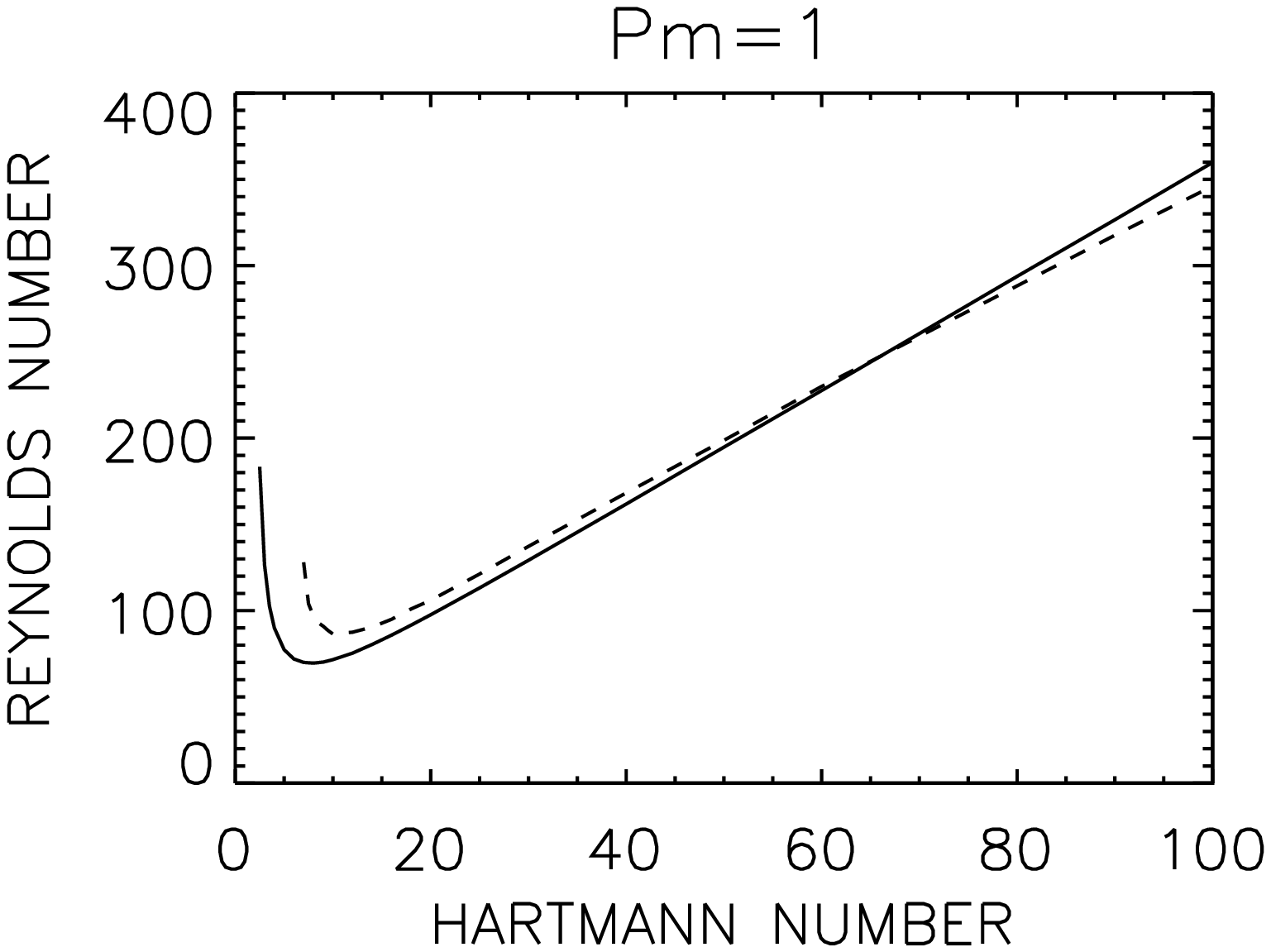,width=4cm,height=8cm}}
\hbox{
\psfig{figure=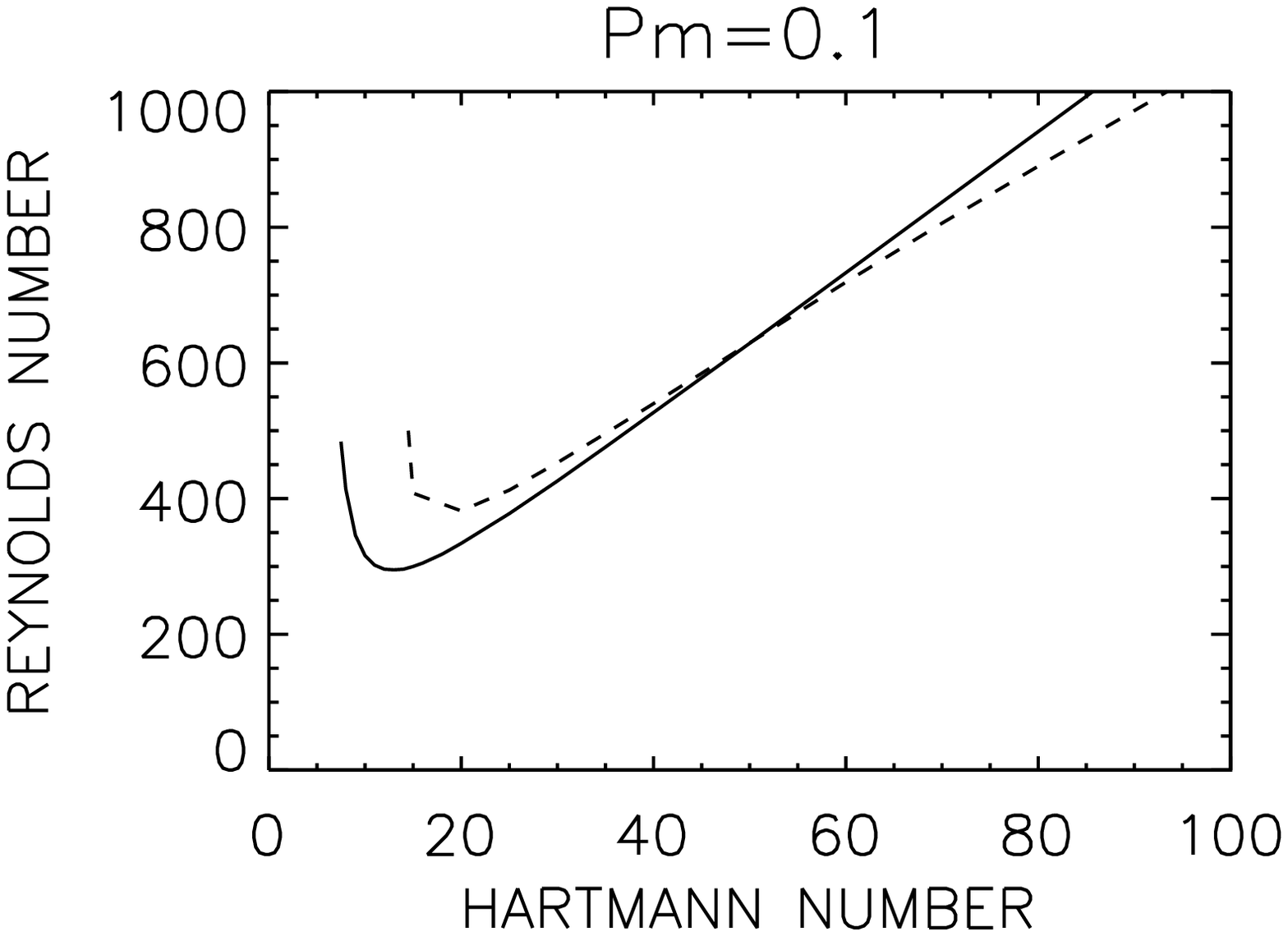,width=4cm,height=8cm}\hfill
\psfig{figure=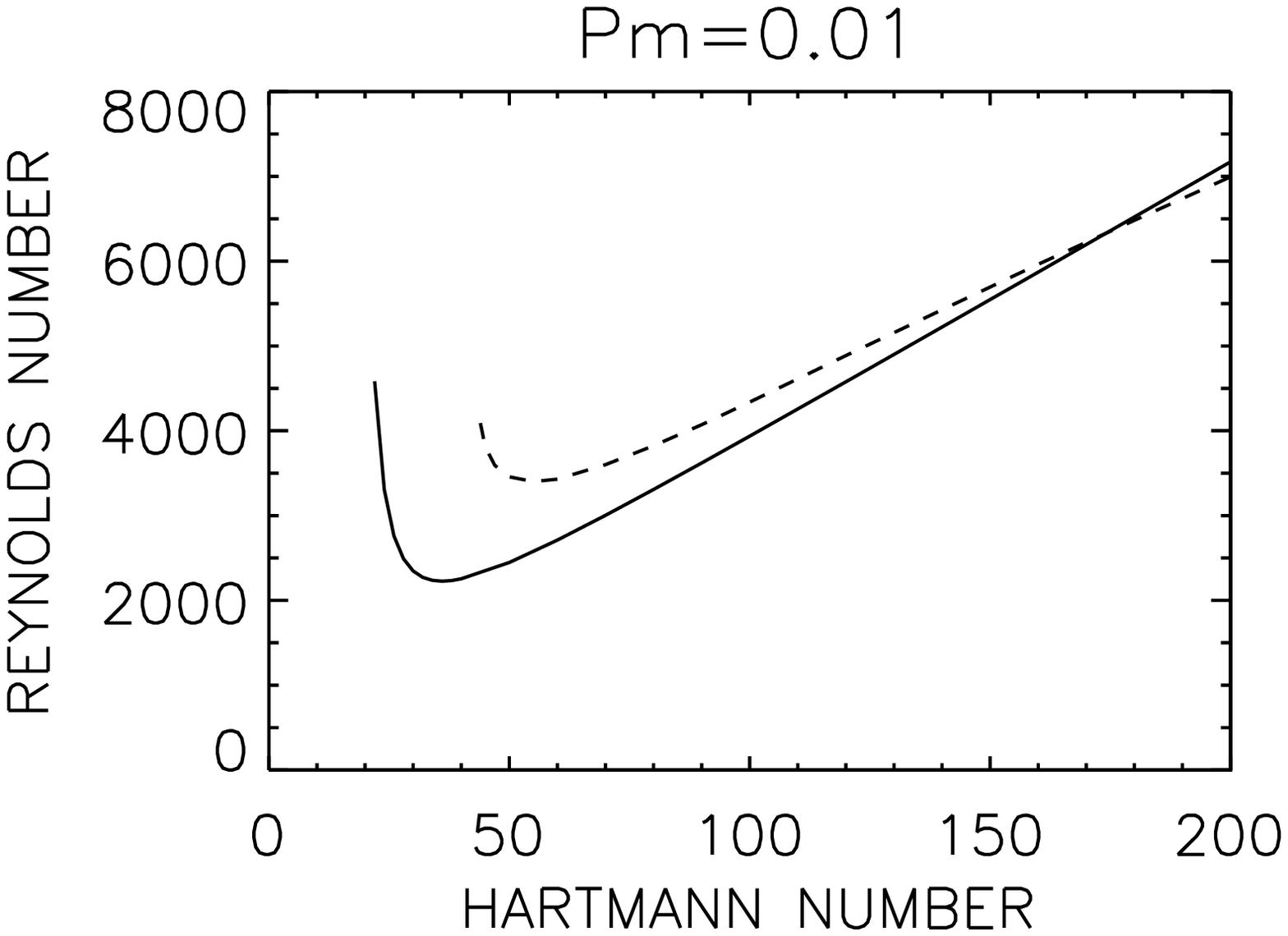,width=4cm,height=8cm}}}
\caption{\label{f2} The same as in Fig. \ref{f1} but for rotating outer 
cylinder ($\hat\mu= 0.33$), there is no hydrodynamical instability}
\end{figure}

If the outer cylinder rotates with an angular velocity $\hat \mu \geq \hat
\eta^2$ than the linear instability without magnetic field disappears and the
critical Reynolds number for Ha $=0$ moves to infinity. However,
for finite Hartmann number (again of order 10) the instability survives 
practically for the same Reynolds numbers. The consequence is the occurrence of typical minima in the stability diagram (Fig. \ref{f2} for$\hat \mu =0.33$).  

The minima also occur for the nonaxisymmetric solutions with $m=1$.  For very high electrical conductivity (Pm=10)  there seems to be no intersection between both the bifurcation profiles. The ring-like structure with $m=0$ always possesses the lowest critical Reynolds number. 

This is not true, however, for smaller magnetic Prandtl numbers, i.e. for lower electrical conductivity. For Pm$\lsim 1$ we always find intersections between the lines for $m=0$ and $m=1$. Again there is a critical Hartmann number
at which the ring geometry ($m=0$) of the excited flow and field pattern changes
to a nonaxisymmetric  geometry  with $m=1$.\footnote{For differentially rotating spheres
we find a similar behavior but only for stress-free boundary conditions (\K\ \&
\R\ 1997)} 

Hence, also in  experiments  with rotating outer cylinder the magnetic field is able to produce nonaxisymmetric structures.
After
the Cowling theorem which requires the existence of nonaxisymmetric
magnetic modes for the existence of a dynamo, a selfexcited dynamo might thus exist, but only   for  certain magnetic Prandtl numbers, i.e. for Pm$\lsim 1$. The magnetic Prandtl number for experiments with liquid metals like sodium or gallium with Pm of order $10^{-(5...6)}$ are still smaller than the considered values.

\section{Wave number and drift frequencies}

\begin{figure}
\hbox{
\psfig{figure=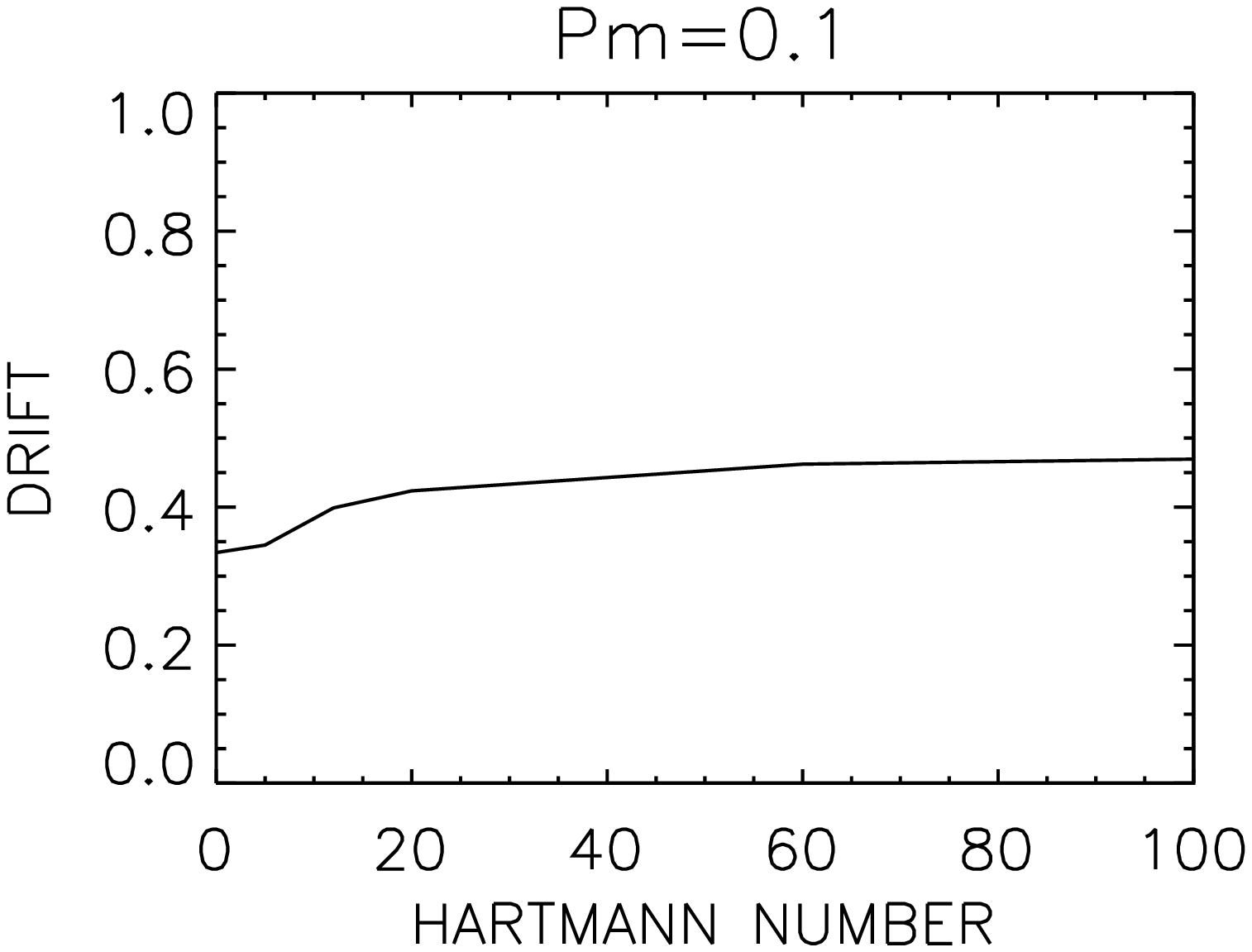,width=4.5cm,height=4.5cm}\hfill
\psfig{figure=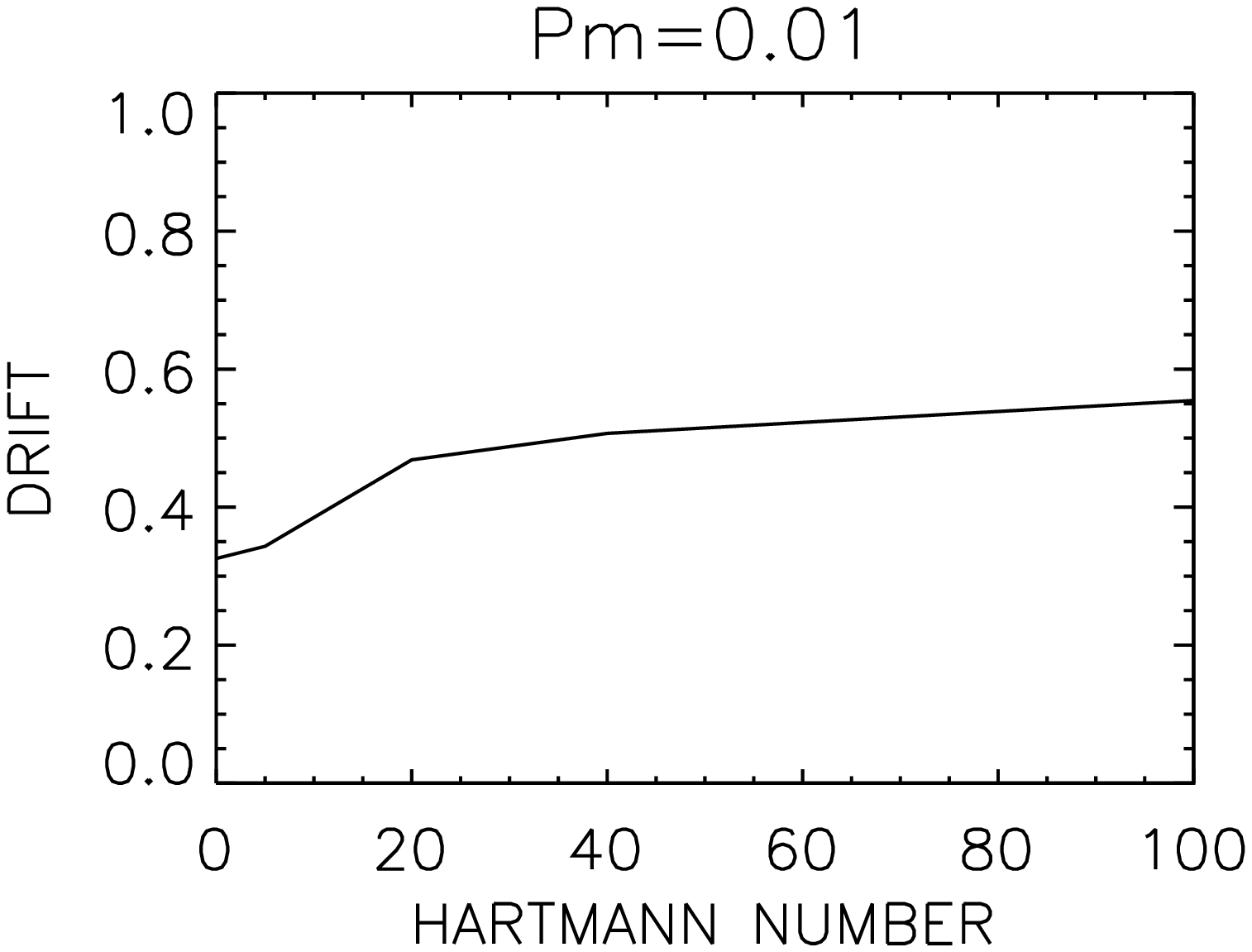,width=4.5cm,height=4.5cm}
}
\caption{\label{drift} The drift frequencies (\ref{dotfi})  normalized with the rotation rate $\Om_{\rm in}$ for the solutions with $m=1$ for resting outer cylinders. The drift of the spirals is always positive, i.e. in the direction of the rotation with about 50 \% of the rotation of the inner cylinder}
\end{figure}

The wave numbers have been discussed in detail for the axisymmetric modes in a
foregoing paper (\R\ \& Shalybkov 2002). Generally,  the cells for the
nonaxisymmetric modes become more and more elongated in the vertical direction. Here we only
add remarks about the drift velocity ${\cal R} (\omega)$ which always proved
to be positive, i.e. the pattern drifts are in direction of the rotation (eastward). It is
\beg
\dot \phi = {{\cal R}(\omega) \Om_{\rm in} \over m {\rm Re}},
\label{dotfi}
\ende
so that for $m=1$ the drift period in units of the rotation period runs as
Re/${\cal R}(\omega)$. A typical value for this ratio is 2.  In Fig. \ref{drift} the expression (\ref{dotfi}) is given normalized with the rotation rate $\Om_{\rm in}$ for the solutions with $m=1$ 
and for the two low magnetic Prandtl numbers that we have considered. 

\section{Conclusions}
We have shown that  a Taylor-Couette flow which is stable in the 
hydrodynamic regime ($\hat \mu  \geq \hat \eta^2$) is destabilized by a 
weak axial magnetic field. Below a critical Hartmann number of order 10... 100  the instability 
sets in in form of axisymmetric rolls while above this value the instability  forms nonaxisymmetric field and flow modes. This phenomenon 
exists despite of the observation (e.g. in dynamo theory) that differential rotation is known as suppressing the 
formation of nonaxisymmetric magnetic fields.  

On the other hand, after the Cowling theorem of dynamo theory a magnetic field  can only be maintained if it is nonaxisymmetric. 
 Considering a  number of typical magnetic Prandtl numbers we find  that for our container with conducting cylinders the dominance of the nonaxisymmetric modes only occurs  for not too high and not too low magnetic Prandtl number. Obviously, the dissipation processes are more important for nonaxisymmetric modes rather than axisymmetric modes. Hence the dissipation  allows  nonaxisymmetric modes only to be preferred if both the dissipation values have nearly the same order of magnitude.


%
\end{document}